\documentclass[a4paper,11pt]{article}
\usepackage{pos}
\usepackage{soul}

\title{Astro-COLIBRI: a new platform for real-time multi-messenger astrophysics}
 \ShortTitle{Astro-COLIBRI}

\author*[a]{, Fabian {Schüssler}}
\author[a]{, Atilla Kaan {Alkan}}
\author[a]{, Valentin {Lefranc}}
\author[a,b,c]{Patrick {Reichherzer}}

\affiliation[a]{IRFU, CEA, Université Paris-Saclay, F-91191 Gif-sur-Yvette, France}
\affiliation[b]{Ruhr-Universität Bochum, Universitätsstraße 150, 44801 Bochum, Germany}
\affiliation[c]{Ruhr Astroparticle and Plasma Physics Center, Ruhr-Universität Bochum, 44780 Bochum, Germany}

\emailAdd{astro.colibri@gmail.com}

\newcommand{\ac}{Astro-COLIBRI}

\abstract{Flares of known astronomical sources and new transient phenomena occur on different timescales, from sub-seconds to several days or weeks. The discovery potential of both serendipitous observations and multi-messenger and multi-wavelength follow-up observations could be maximized with a tool which allows for quickly acquiring an overview over both persistent sources as well as transient events in the relevant phase space. We here present COincidence LIBrary for Real-time Inquiry (Astro-COLIBRI), a novel and comprehensive tool for this task.

Astro-COLIBRI's architecture comprises a RESTful API, a real-time database, a cloud-based alert system and a website as well as apps for iOS and Android as clients for users. The structure of Astro-COLIBRI is optimized for performance and reliability and exploits concepts such as multi-index database queries, a global content delivery network (CDN), and direct data streams from the database to the clients to allow for a seamless user experience. Astro-COLIBRI evaluates incoming VOEvent messages of astronomical observations in real time, filters them by user-specified criteria and puts them into their MWL and MM context. The clients provide a graphical representation with an easy to grasp summary of the relevant data to allow for the fast identification of interesting phenomena and provides an assessment of observing conditions at a large selection of observatories around the world.

We here summarize the key features of Astro-COLIBRI, the architecture and used data resources. We specifically provide examples for applications and use cases. Focussing on the high-energy domain, we showcase how Astro-COLIBRI facilitates the search for high-energy gamma-ray counterparts to high-energy neutrinos and scheduling of follow-up observations of a large variety of transient phenomena like gamma-ray bursts, gravitational waves, TDEs, FRBs, and others.}

\FullConference{37$^{\rm{th}}$ International Cosmic Ray Conference (ICRC 2021)\\
		July 12th -- 23rd, 2021\\
		Online -- Berlin, Germany}

\begin{document}
\maketitle

\section{Introduction}
Time-domain astrophysics, the study of transient phenomena is a recent and rapidly increasing field of astronomy. We here use a very broad definition of {\it transient phenomena} referring to astrophysical sources that show an un-predictable temporal (and mostly also) spatial behaviour. Focussing on sources of interest for observatories in the high and very-high-energy domain, these are thus outbursts, flares or explosions of previously known objects and include a large variety of objects like solar or other stellar flares, novae and supernovae (SNe), (giant) flares from pulsars, bursts of soft gamma-ray repeaters (SGRs), anomalous X-ray pulsars (AXPs), changes in the emission of X-ray binaries, microquasars, and flares of active galactic nuclei (AGN).

A second, and not always clearly separated, class of transient events are those that can appear at any point in the sky without an a priori clear association to an existing astrophysical object. These include Gamma-Ray Bursts (GRBs), Fast Radio Bursts (FRBs), tidal disruption of a stellar object in the vicinity of a supermassive black hole (tidal disruption events, TDE), mergers of a massive binary system detected in gravitational waves (GW), or the detection of one or several high-energy neutrinos. As seen in the last points, the transient phenomena can also be defined solely by the detection of particular events that may indicate transient behaviour of an astrophysical source. 

Most if not all of these phenomena are detected by either, large field-of-view monitoring instruments (e.g. Fermi-GBM and Swift-BAT for GRBs, Fermi-LAT for AGN flares, VIRGO/Ligo/KAGRA for GWs, CHIME for FRBs, IceCube for neutrinos, etc.). Subsequent to the initial detection of transient phenomena, and despite the large variety of events, dedicated follow-up observations are typically necessary to confirm the detection, obtain higher (temporal, spatial, energy, etc.) resolution observations necessary for the characterisation and detailed understanding of the underlying processes. These follow-up observations are in general performed by pointing instruments that are operated via competitive call for proposals. The decision to request observations for a particular (transient) event has therefore to be based on the most complete, up-to-date and precise information available. Providing this information to the community in an accessible way, able to keep up the pace with the ever-increasing amount of information provided by the more and more sensitive observatories is the main purpose of the Astro-COLIBRI platform.

\section{The Astro-COLIBRI platform}
\begin{figure}[t!]
\centering
\includegraphics[width=1.0\linewidth]{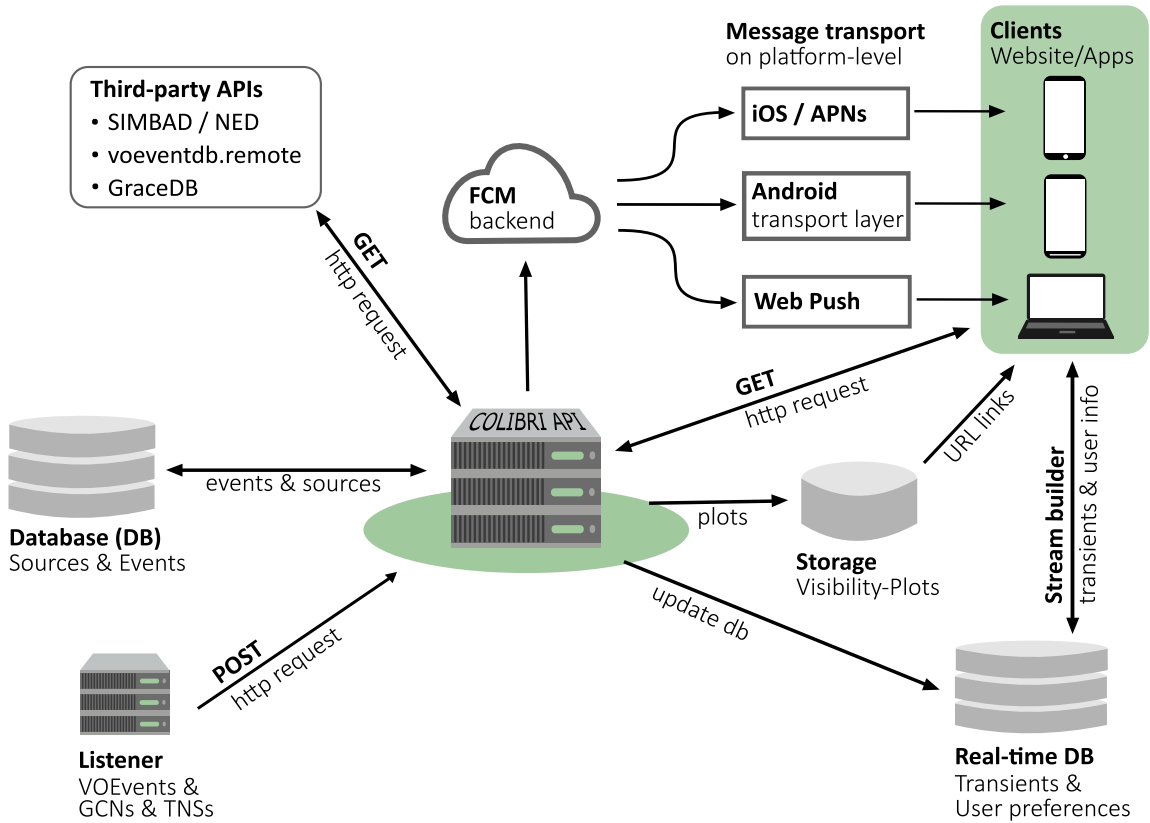}
\caption{Architecture of Astro-COLIBRI composed of different interconnected modules. The fields highlighted in green are public interfaces.  \label{fig:architecture}}
\end{figure}

A large amount of information is available on known astrophysical objects and on transient phenomena. Often due to historical reasons, this wealth of information is unfortunately spread over many systems, databases, and collaborations. For example, SIMBAD~\citep{Simbad_2000}, NED~\citep{NED_1991}, etc. provide detailed information about known sources and the associated archival observations. The Gamma-Ray Coordinates Network (GCN~\citep{2000AIPC..526..731B, GCN}) is providing an efficient real-time stream of announcements of mainly GRBs, GW, and high-energy neutrinos, while the Transient Name Server (TNS~\citep{TNS}) is providing similar information for SNe, TDEs, FRBs, etc. This is complemented by collaborations operating monitoring instruments like Fermi-LAT that provide analyses of their data at various timescale (e.g. the Fermi All-sky Variability Analysis (FAVA~\citep{FAVA_2017})) and community platforms like the Astronomers Telegram (ATEL~\citep{ATEL}) enable sharing of reports on observations.

This huge, and increasing, amount of information can be used to efficiently select the most promising candidate events for scheduling of deeper follow-up observations, but keeping track of these various sources of information in real-time (and partially in stressful situations while operating an instrument in the middle of the night) can be time-consuming and error-prone. Without replacing the individual, expert-run systems, Astro-COLIBRI provides a top-level layer enabling access to the most important information at a glance. 

To fulfil this promise, Astro-COLIBRI is built as a modular, cloud-based infrastructure that is using up-to-date technologies like a central RESTful API, a real-time database, a cloud-based alert system and a website as well as apps for iOS and Android as clients for users. The structure of Astro-COLIBRI (cf. Fig.~\ref{fig:architecture}) is optimized for performance and reliability and exploits concepts such as multi-index database queries, a global content delivery network (CDN), and direct data streams from the database to the clients to allow for a seamless user experience. Dedicated listeners are subscribed to the most relevant alert networks using the IVAO~\citep{IVOA} standard protocols and each incoming VOEvent message is analyzed, filtered, and classified in real-time. The API is using dedicated, cloud-based databases and access to external APIs to put the incoming alert into its multi-wavelength (MWL) and multi-messenger (MM) context. 

The deeply customizable clients provide a graphical representation with an easy to grasp summary of the relevant data to allow for the fast identification of interesting phenomena and provide an assessment of observing conditions at a large selection of observatories around the world.

Astro-COLIBRI is currently approaching the end of an extensive beta-testing phase. A fully functional setup of the outlined architecture has been deployed and tested over several months. We achieved high reliability, an almost $100~\%$ uptime and fast processing speeds. Incoming alerts are announced to subscribed mobile devices well under one second and most queries to the API are answered within few seconds, sufficient for an optimal user experience. 

Publication of the developed Android and iOS applications in the relevant app stores is imminent. Online tutorials and hands-on demonstrations of the main features are being compiled on a dedicated YouTube channel~\cite{YouTube}, while the public endpoints of the API are fully documented~\cite{API_docu}. 
\begin{figure}[t!]
\centering
\includegraphics[width=1.0\linewidth]{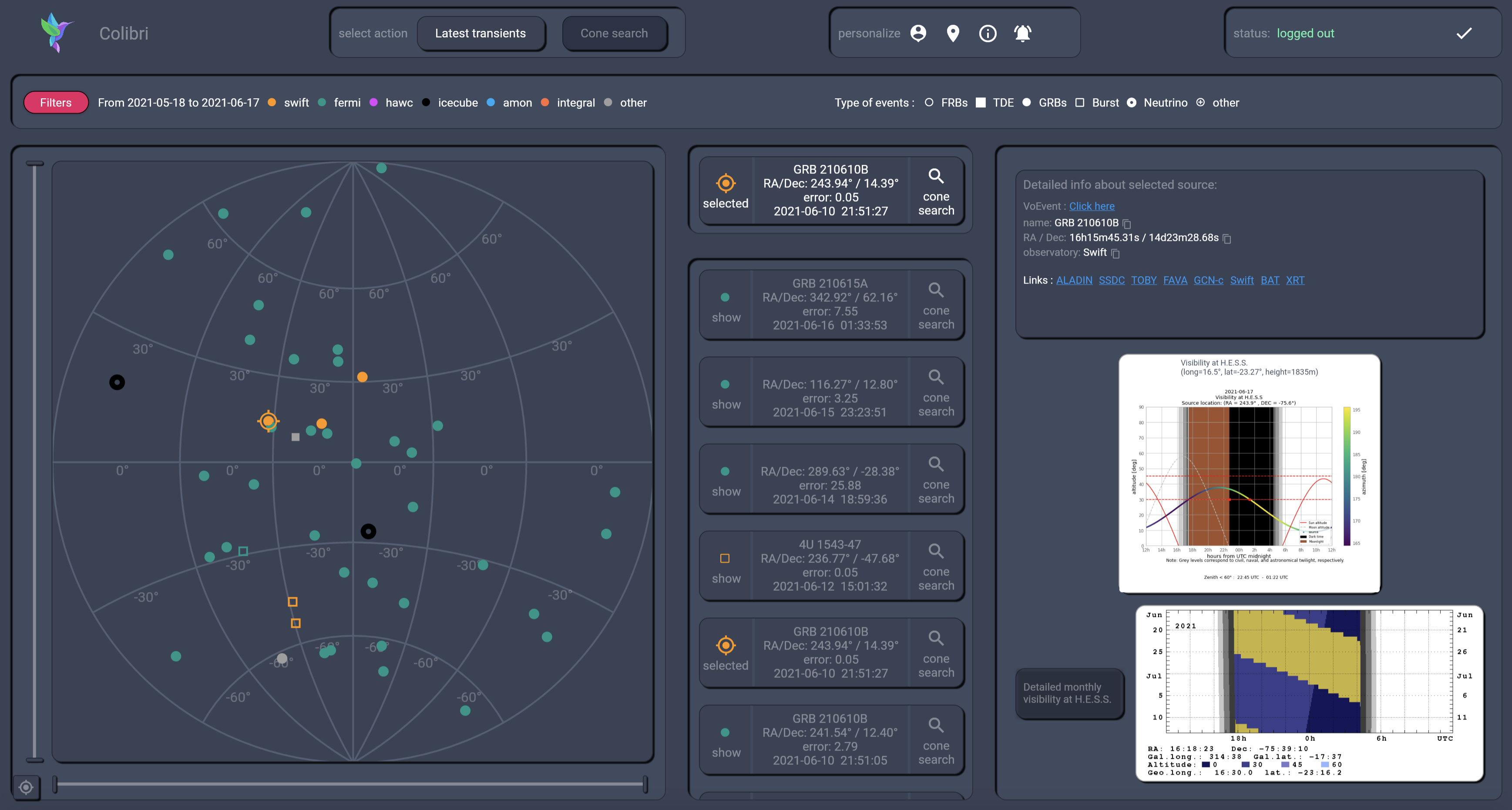}
\caption{The Astro-COLIBRI web-based interface showing by default transient events detected over the last month in a sky-view and a time-ordered list. Both allow to select events of interest. Relevant information, links to external services and the visibility at a given observatory are shown in the right panel. \label{fig:web}}
\end{figure}

\section{Example use-case: observations of GRBs}
One of the most common transient events of interest for the high-energy astrophysics community are GRBs. Providing information on their detections and enabling well-informed decision to trigger follow-up observations with ground or space-based observatories is therefore obviously a core feature of \ac{}. A dedicated server with extensive performance monitoring is constantly listening to a redundant list of IVOA VoEvent brokers. Upon the announcement of a new GRB by any of the most active X-ray monitors (Fermi-GBM, Swift-BAT, INTEGRAL), the \ac{} pipeline outlined in Fig.~\ref{fig:architecture} is activated. The (python based) logic in the API searches the \ac{} databases for existing entries of the same event and classifies the incoming information both following the expected sequence of messages from each observatory as well as using the data provided in the alert itself (e.g. the GCN \verb!Pkt_Ser_Num!, references to previous messages on the same event, etc.). If appropriate, the databases are updated to always contain the newest and highest precision information on each event. Thanks to the established direct data streams between the real-time database and the clients (both web and smartphone apps), the latter are updated automatically with negligible delays. The name of a new GRB (e.g. \verb!GRB YYMMDDA/B/.!) is typically only communicated by the detecting observatories via GCN circulars. Dedicated listeners parsing the associated email messages allow adding this information fully automatically. While the announcement of the detection of GeV emission by Fermi-LAT is also being treated automatically, further information can easily be added manually by the \ac{} admins.

\begin{figure}[t!]
\centering
\includegraphics[width=1.0\linewidth]{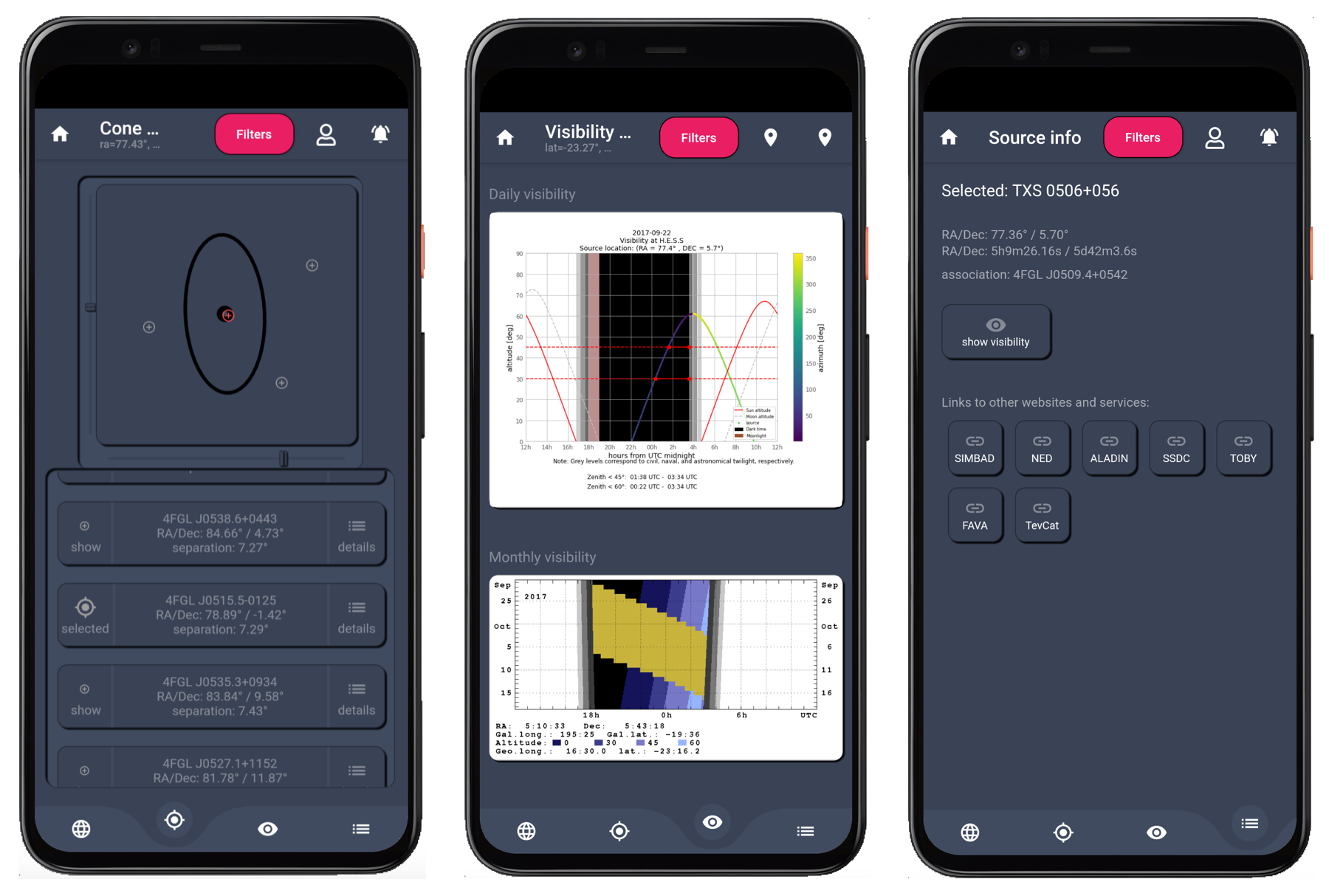}
\caption{Example of the \ac{} smartphone application providing access to information and context related to transient phenomena like high-energy neutrinos (here: IceCube-170922A / TXS 0506+056).\\[0.5ex] 
Left panel: Result of the cone search around the neutrino event showing known GeV and TeV emitters.\\[0.5ex] 
Middle panel: Visibility of the source today (e.g. 2017-09-22 for H.E.S.S.) and over the next month\\[0.5ex] 
Right panel: Detailed information about the event and access to external information. \label{fig:mobile_txs0506}}
\end{figure}

The client interfaces summarize the collected information in a compact display (cf. the upper right box in Fig.~\ref{fig:web} and the right panel in Fig.~\ref{fig:mobile_txs0506}). It includes direct access to information from SIMBAD, NED, ALADIN~\citep{Aladin_2014}, a MWL SED compiled by the SSDC~\citep{SSDC}, the recent GeV activity as seen by Fermi-LAT via FAVA, links to the full history of GCN notices and circulars, as well as access to the analysis results of the different instruments (e.g. Swift-BAT, Swift-XRT, Fermi-GBM, etc.). Following suggestions by the growing community of \ac{} users, additional information is being made added regularly.     

The provided information is completed with an assessment of the visibility of the selected event at a large variety of observatories. The visibility and scheduled observations of all major space-born observatories (incl. INTEGRAL, Chandra, NuSTAR, XMM-Newton, Insight-HXMT, Swift, etC.) is accessible via a direct link to the IVAO prototype tool TOBY. Ground based observatories have typically more stringent constraints on the observability of a given direction at a given time. These are summarized in two displays that are customizable via the selection of an observatory: a first figure is providing a detailed assessment of the observability of the selected event during the current day taking into account constraints imposed by the sun as well as the moon. Complementary information is shown in a second figure that illustrates the longer-term visibility over the upcoming month.

\section{Example use-case: multi-messenger searches for the sources of high-energy neutrinos}
Searches for (transient) electromagnetic emission associated to high-energy neutrinos is one of the most promising avenues to pin-point the acceleration sites of very-high and ultra-high energy cosmic rays. For several years, IceCube, the leading neutrino telescope, is publicly announcing the detection of potentially interesting neutrino candidates to the community. The significant efforts in conducting follow-up observations of these alerts have seen a first success in the identification of a flaring state of a blazar emitting up to the VHE gamma-ray domain, TXS 0506+056, possibly related to the high-energy neutrino named IC-170922A~\citep{2018Sci...361.1378I}. Despite significant interest and a global campaign of observations, it took roughly one week between the detection of the neutrino event and the realization that there is a blazar that had been in an elevated state in the GeV domain for several months located within the error box of the neutrino~\citep{ATEL_TXS_flare}. While all necessary data for this crucial assessment, which then triggered a global observation campaign, was readily available via various public catalogs and services like FAVA, these checks and searches had to be done manually, thus inducing unnecessary delays.

We believe that it is therefore of utmost importance to provide a single, complete and easy to use access point to a large set of information. All public alert streams from neutrino telescopes are included in the \ac{} pipeline. They are announced via a dedicated notification stream, the user can subscribe to on both Android and iOS mobile platforms. These notifications already contain a summary of the most crucial information like the event coordinates and their uncertainty. 

For this specific science case, the automatic cone search provided by \ac{} is a central element. Displayed via the web and smartphone frontends, it immediately provides a visual indication of previous transients and known GeV gamma-ray emitters within the neutrino uncertainty circle and the region around that. The provided card-view also includes the angular distance to the neutrino event to allow for a first quantitative assessment. The cone search result is also rapidly available in a machine-readable format via the dedicated API endpoint \verb!/cone_search!~\cite{API_docu}. 

While the visibility plots allow assessing the observability of the given event by the user-selected observatory, the information box provides a detailed overview of the event information like the false-alarm rate (FAR) and the probability of the event to be of astrophysical origin ($P_\mathrm{astro}$). Further information can be found in the original VoEvent file announcing the event that is provided via a call to the 4PiSky database and via links to collections of GCN notices and circulars related to the event. Additional links to all relevant external services enable a quick assessment of the event and possible counterparts. These include direct access to information from the general-purpose services SIMBAD, NED, ALADIN, SSDC, FAVA and archival TeV gamma-ray information from TeVCat. \ac{} is thus providing a central platform that allows accessing all necessary information for a rapid and informed case-by-case decision on the interest of a particular neutrino alert.

\section{Outlook}
During the beta testing phase, the focus was put on the development and improvement of the user interfaces and frontend clients. With the end of the beta phase, this will shift towards deployment of \ac{} within various observatories. Several levels of integration are possible. They range from the use of the \ac{} clients by individual burst advocates to the deployment of the (web) client in observatory control rooms and even the submission of (possibly confidential) alerts created by real-time data analysis frameworks into the \ac{} pipeline. The first step into this latter direction is already implemented: members of the H.E.S.S. collaboration have access to information derived from a dedicated analysis of Fermi-LAT data via FLaapLUC~\citep{FlaapLUC2018}. Similar partnerships are under discussion with other collaborations. 

In order to allow other observatories to conduct follow-up observations, the information related to the detection of a new transient phenomena has to be communicated very rapidly within the astrophysicist community. While a part of these information are shared in machine readable format (VOEvent messages), a significant amount of data, especially about results from follow-up observations, are disseminated via manual reports written by humans (e.g. GCN circulars, ATELs, etc.). The improvement of observation techniques and the increased interest in time domain astrophysics has resulted in a dramatic increase in the number of these reports. This leads to a saturation of the way astrophysicists read, analyze and classify information, calling for new ways of handling and extracting information in text files. That is the reason why, as further functionalities for the \ac{} platform, we are aiming to build and develop models based on Natural Language Processing (NLP) techniques to classify and extract information from observation reports distributed in natural language. Such information extraction models have already been developed in other specialized domains such as clinical reports, and first results using deep neural networks are promising~\citep{biobert, nlp_ner}. In collaboration with NLP experts, the longterm goal is to be able to automatically analyze and summarize human written observation reports in real-time, thus potentially providing a significant improvement to the \ac{} platform. 

The \ac{} development team is welcoming comments and feedback from the community to further improve the platform. Contact: \href{mailto:astro.colibri@gmail.com}{astro.colibri@gmail.com}

\acknowledgments
This work is supported by the European Union’s Horizon 2020 Program under the AHEAD2020 project (grant agreement n. 871158).

\bibliographystyle{ICRC}
\bibliography{references}

\providecommand{\href}[2]{#2}\begingroup\raggedright\begin{thebibliography}{10}

\bibitem{Simbad_2000}
M.~{Wenger} {\em et~al.}, \href{http://dx.doi.org/10.1051/aas:2000332}{``{The
  SIMBAD astronomical database. The CDS reference database for astronomical
  objects},''} in {\em A\&A, Suppl.}, vol.~143, pp.~9--22.
\newblock Apr., 2000.

\bibitem{NED_1991}
G.~Helou {\em et~al.}, {\em The NASA/IPAC Extragalactic Database},
  \href{http://dx.doi.org/10.1007/978-94-011-3250-3_10}{pp.~89--106}.
\newblock Springer Netherlands, Dordrecht, 1991.

\bibitem{2000AIPC..526..731B}
S.~D. {Barthelmy} and others.,
  \href{http://dx.doi.org/10.1063/1.1361631}{``{GRB Coordinates Network (GCN):
  A status report},''} in {\em Gamma-ray Bursts, 5th Huntsville Symposium},
  vol.~526 of {\em AIP Conf. Series}, pp.~731--735.
\newblock Sept., 2000.

\bibitem{GCN}
``{The Gamma-ray Coordinates Network}.'' \url{https://gcn.gsfc.nasa.gov/}.

\bibitem{TNS}
``{Transient Name Server}.'' \url{https://www.wis-tns.org/}.

\bibitem{FAVA_2017}
S.~{Abdollahi} {\em et~al.},
  \href{http://dx.doi.org/10.3847/1538-4357/aa8092}{``{The Second Catalog of
  Flaring Gamma-Ray Sources from the Fermi All-sky Variability Analysis},''} in
  {\em APJ}, vol.~846, p.~34.
\newblock Sept., 2017.

\bibitem{ATEL}
``{The Astronomer's Telegram}.'' \url{https://www.astronomerstelegram.org/}.

\bibitem{IVOA}
``{International Virtual Observatory Alliance}.'' \url{https://www.ivoa.net/}.

\bibitem{YouTube}
``{Astro-COLIBRI YouTube channel}.''
  \url{https://www.youtube.com/channel/UClDhbN-S_nMPLoVxLF6ouIQ}.

\bibitem{API_docu}
``{Astro-COLIBRI API documentation}.''
  \url{https://astro-colibri.herokuapp.com/documentation}.

\bibitem{Aladin_2014}
T.~{Boch} and P.~{Fernique}, ``{Aladin Lite: Embed your Sky in the Browser},''
  in {\em Astronomical Data Analysis Software and Systems XXIII}, N.~{Manset}
  and P.~{Forshay}, eds., vol.~485 of {\em ASP Conf. Series}, p.~277.
\newblock May, 2014.

\bibitem{SSDC}
``{Space Science Data Center}.'' \url{https://tools.ssdc.asi.it/SED/}.

\bibitem{2018Sci...361.1378I}
{The IceCube, Fermi-LAT, MAGIC, AGILE, ASAS-SN, HAWC, H.E.S.S, INTEGRAL,
  Kanata, Kiso, Kapteyn, Liverpool telescope, Subaru, Swift/NuSTAR, VERITAS,
  VLA/17B-403 teams} \href{http://dx.doi.org/10.1126/science.aat1378}{{\em
  Science} {\bfseries 361} no.~6398, (July, 2018) eaat1378}.

\bibitem{ATEL_TXS_flare}
``{Fermi-LAT detection of increased gamma-ray activity of TXS 0506+056, located
  inside the IceCube-170922A error region}.''
  \url{https://www.astronomerstelegram.org/?read=10791}.

\bibitem{FlaapLUC2018}
J.~P. {Lenain},
  \href{http://dx.doi.org/10.1016/j.ascom.2017.11.002}{``{FLaapLUC: A pipeline
  for the generation of prompt alerts on transient Fermi-LAT
  {\ensuremath{\gamma}}-ray sources},''} in {\em Astronomy and Computing},
  vol.~22, pp.~9--15.
\newblock Jan., 2018.

\bibitem{biobert}
J.~Lee {\em et~al.}, ``{BioBERT: a pre-trained biomedical language
  representation model for biomedical text mining},'' in {\em Bioinformatics}.
\newblock Oxford University Press (OUP), Sep, 2019.

\bibitem{nlp_ner}
Z.~Huang {\em et~al.}, ``{Bidirectional LSTM-CRF Models for Sequence
  Tagging},'' in {\em arXiv}.
\newblock 2015.
\newblock \href{http://arxiv.org/abs/1508.01991}{{\ttfamily arXiv:1508.01991
  [cs.CL]}}.

\end{thebibliography}\endgroup

\end{document}